\documentclass[12pt]{iopart}
\usepackage{cite}
\usepackage{graphicx}%
\usepackage{hyperref}
\usepackage[dvipsnames]{xcolor}
\usepackage{ulem}

\def\comAK#1{\textcolor{black}{#1}}
\def\comAB#1{\textcolor{black}{#1}}

\begin{document}

%\title{Electronic correlation effects and magnetic properties of L1$_0$ phase of FeNi}

\title{Electronic correlation effects and local magnetic moments in L1$_0$ phase of FeNi}

\author{A S Belozerov$^{1}$, A A Katanin$^{2,1}$, V I Anisimov$^{1,3,4}$}

\address{$^1$ M. N. Miheev Institute of Metal Physics, Russian Academy of Sciences, 620108 Yekaterinburg, Russia}
\address{$^2$ Moscow Institute of Physics and Technology, 141701 Dolgoprudny, Russia}
\address{$^3$ Skolkovo Institute of Science and Technology, 121205 Moscow, Russia}
\address{$^4$ Ural Federal University, 620002 Yekaterinburg, Russia}

\ead{alexander.s.belozerov@gmail.com}

%\author{A. S. Belozerov}
%\affiliation{Skolkovo Institute of Science and Technology, 121205 Moscow, Russia}
%\affiliation{M. N. Miheev Institute of Metal Physics, Russian Academy of Sciences, 620108 Yekaterinburg, Russia}

%\author{A. A. Katanin}
%\affiliation{Moscow Institute of Physics and Technology, 141701 Dolgoprudny, Russia}
%\affiliation{M. N. Miheev Institute of Metal Physics, Russian Academy of Sciences, 620108 Yekaterinburg, Russia}
%\affiliation{Skolkovo Institute of Science and Technology, 121205 Moscow, Russia}

%\author{V. I. Anisimov}
%\affiliation{Skolkovo Institute of Science and Technology, 121205 Moscow, Russia}
%\affiliation{M. N. Miheev Institute of Metal Physics, Russian Academy of Sciences, 620108 Yekaterinburg, Russia}
%\affiliation{Ural Federal University, 620002 Yekaterinburg, Russia}

\begin{abstract}
We study the electronic and magnetic properties of L1$_0$ phase of FeNi, a perspective rare-earth-free permanent magnet, by using a combination of density functional and dynamical mean-field theory. Although L1$_0$ FeNi has a slightly tetragonally distorted fcc lattice, we find that magnetic properties of its constituent Fe atoms resemble those in pure bcc Fe. In particular, our results indicate the presence of well-localized magnetic moments on Fe sites, which are formed due to Hund's exchange. At the same time, magnetism of Ni sites is much more itinerant. Similarly to pure bcc Fe, the self-energy of Fe $3d$ states is found to show the non-Fermi-liquid behavior. This can be explained by peculiarities of density of Fe $3d$ states, which has pronounced peaks near the Fermi level. Our study of local spin correlation function and momentum dependence of particle-hole bubble suggests that the  magnetic exchange in this substance is expected to be of RKKY-type, with iron states providing local-moment contribution, and the states corresponding to nickel sites (including virtual hopping to iron sites) providing itinerant contribution.
\end{abstract}
%\submitto{\JPCM}

\maketitle

\section{Introduction}

Growing demand in recent decades for high-performance permanent magnets  and deficiency of rare-earth elements 
%containing in them 
have stimulated an active search of new rare-earth-free magnets. 
%magnets, which are free of REE.
%
One of the promising materials is ordered L1$_0$ phase of FeNi, %, also called tetrataenite.
which was first discovered in 1962 during the neutron irradiation of disordered FeNi alloy~\cite{Pauleve1962}. 
Later, samples of L1$_0$ FeNi were found in meteorites, and
%in 1980
this phase was named tetrataenite after its tetragonal crystal structure and \mbox{Fe-Ni} alloy taenite,   
also 
found in meteorites.
Formation of tetrataenite in meteorites occurs due to extremely slow cooling ($\sim$0.1~K per million years) at a temperature of about 
%593
600~K.
%\comAK{Above this temperature the tetrataenite becomes structurally unstable towards disordered phase?}
%\comAB{(AB: Probably, a similar discussion is more suitable just before/in the last paragraph of Introduction.)}\comAK{I simply wanted to mention this here, as an experimental fact}.
%\comAB{(Alternative version: ok
\comAB{
%Upon further heating
Above this temperature the tetrataenite becomes metastable up to the Curie temperature of 823~K\cite{wasilewski1988,Lewis2014}.}
Therefore, an artificial synthesis of its significant amount is challenging.
%
%Nevertheless, a
In this respect, a successful fabrication of well-ordered film~\cite{Shima2007,Kojima2014}, bulk~\cite{Makino2015} and single-phase powder samples~\cite{Goto2017} was recently reported.

Tetrataenite has been extensively studied experimentally in the meteorite~\cite{Kotsugi2014,Lewis2014,wasilewski1988,Poirier2015_Bordeaux2016} and artificially synthesized~\cite{Pauleve1962,Shima2007,Kojima2014,Makino2015,Pauleve1968,Neel1964_Chamberod1979_Reuter1989_Lee2014,Mizuguchi2011,Frisk2016} samples,
that helped %in understanding of the magnetic properties~\cite{Kotsugi2014,Lewis2014}, and clarified their dependence 
to clarify 
%the dependence of 
its magnetic properties 
%on the long-range order~\cite{Kojima2014} 
and their dependence on structural parameters~\cite{Shima2007,Mizuguchi2011}.
In contrast to the disordered Fe-Ni alloys, tetrataenite can be classified as a hard magnet due to large coercivity of more than 500~Oe~\cite{Kotsugi2014,Lewis2014}.
Other characteristics of tetrataenite are also comparable to those of rare-earth-based magnets.
In particular, it has
%a large crystalline anisotropy of 1.1-1.3~MJ/m$^3$~\cite{},
%a large coercivity of $\sim$500~Oe~\cite{kotsugi2014},
a high Curie temperature of 823~K~\cite{wasilewski1988},
a saturation magnetization of $\sim$1270~emu/cm$^3$~\cite{Pauleve1968},
and the theoretical magnetic energy product of $\sim$42~MG Oe~\cite{Lewis2014}, which is close to that of Sm-Co and Nd-Fe-B magnets.

\begin{figure}[b]
\centering
\includegraphics[clip=true,width=0.29\textwidth]{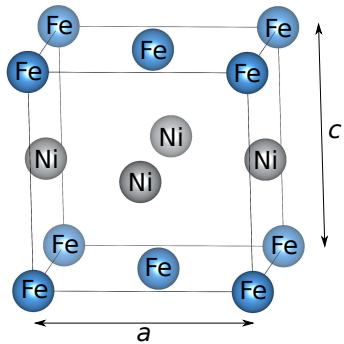}
\caption{%(Color online)
\label{fig:structure}
Crystal structure of the L1$_0$ phase of FeNi. %The iron (nickel) atoms are depicted by blue (green) balls.
The figure was prepared with the VESTA program~\cite{vesta}.}
\end{figure}

The tetragonal L1$_0$ structure of FeNi is of AuCu-type and 
%associated with 
characterized by chemical ordering, in which Fe and Ni monolayers are alternating along the $c$-axis (see Fig.~\ref{fig:structure}).
%
%Tetrataenite has a large crystalline anisotropy (1.1-1.3~MJ/m$^3$)
%caused by tetragonal distortion associated with the chemical ordering.
%
Although tetrataenite does not contain heavy elements, such as $4d$
or $5d$ transition metals, the tetragonal distortion 
%provides 
serves as a source of large magnetocrystalline anisotropy~\cite{Pauleve1968}.
%of $\sim$1.3~MJ/m$^3$~\cite{}.
%
The distortion ${c/a=1.007}$ in terms of face-centered unit cell parameters is however small; 
%
%This implies that the lattice of ordered FeNi is close to 
%
%At the same time, 
a transformation to body-centered cubic (bcc) lattice requires much smaller 
%distortion leading to 
${c/a=1/\sqrt{2}}$ in the same notations.

Theoretical studies of L1$_0$ FeNi have been performed mainly within density functional theory (DFT).
These studies addressed the origin of 
magneto-crystalline anisotropy~\cite{Miura} and
its dependence
%dependence of magnetic properties
on the % degree of
chemical disorder~\cite{Kota,Izardar},
which was shown to reduce both the Curie temperature and magnetic anisotropy energy (MAE)~\cite{Izardar,Edstrom,Tian}.
The latter appeared to be quite sensitive to the approach employed and computational details~\cite{werwinski2017}.
Interestingly, Izardar \textit{et al} found that certain
configurations with partial disorder have a larger MAE than
the perfectly ordered L1$_0$ structure~\cite{Izardar}.
%
%Nevertheless, results obtained by Izardar \textit{et al} suggest that the MAE can be potentially increased beyond the value obtained for the perfectly ordered structure~\cite{Izardar}.
%
Moreover, the % ordering transformation and
phase stability in FeNi was also considered using DFT calculations~\cite{Mohri,Barabash,Tian,Tian2020}.
%
%obtained the order-disorder transition temperature of 559~K compared to the experimental value of 593~K 
%
%In particular, Tian \textit{et al} obtained an accurate estimate of the order-disorder transition temperature and found its strong dependence on the configurational and vibrational degrees of freedom~\cite{Tian}.
%
In particular, Tian \textit{et al} obtained an accurate estimate of the order-disorder transition temperature $T_{\rm od}$ by taking into account configurational and vibrational degrees of freedom~\cite{Tian}.
The Curie temperature and $T_{\rm od}$
were also found to be affected by lattice expansion~\cite{Tian2020}.

The DFT alone, however, has difficulties in description of electron correlations, which were shown to play an important role in iron~\cite{alpha_iron2010,OurAlphaIgoshevKatanin,OurAlphaBelozerovKatanin,leonov_fe,Katsnelson_Grechnev_Benea2012,Hausoel,Minar2005}, nickel~\cite{Katsnelson_Grechnev_Benea2012,Hausoel,Minar2005}, and their alloys~\cite{Hausoel,Minar2005,Leedahl}.
An accurate treatment of local spin dynamics and many-body effects, which can be especially relevant at finite temperatures, is provided by a combination of DFT and dynamical mean-field theory (DMFT)~\cite{dmft}.
%This combination is called the DFT+DMFT approach and allows one to carefully take into account local quantum fluctuations.
%
%However, previous studies of iron~\cite{}, nickel~\cite{}, and their alloys~\cite{} have shown an important role played by electronic correlations in these substances. 
%
%Recent dynamical mean-field theory (DMFT) study of ordered FeNi 
%
%Recent dynamical mean-field theory (DMFT) study of ordered FeNi 
%Recently the DFT+DMFT approach was 
This combination, called DFT+DMFT~\cite{dftdmft}, was recently applied by Benea \textit{et al}~to describe the Compton profile spectra of ferromagnetic L1$_0$ FeNi~\cite{benea2018}.
%
%In recent study of the Compton profile spectra of ordered FeNi~\cite{benea2018}, the many-body effects were taken into account within dynamical mean-field theory (DMFT)~\cite{dmft}.
%
This study has demonstrated the limitation of local spin-density approximation in the low momentum region, where the local dynamical correlations were found to be essential for description of the experimental spectra.
%
%The dynamical correlations for ordered FeNi were previously taken into account by Benea \textit{et al}~\cite{benea2018}
%
%showed that the discrepancy at
%low momenta due to the inadequate treatment of electronic
%correlations in LSDA can be corrected using DMFT. 
%
These results obtained for ferromagnetic state show the importance of further studying the correlation effects in L1$_0$ FeNi.

%\comAB{(Alternative version of the first two sentences:)}

\comAB{In this paper we explore the electronic and dominating magnetic correlations, as well as the formation of local magnetic moments in tetrataenite by DFT+DMFT approach. To this end, we 
%\comAK{enforce paramagnetic phase below}
artificially switch off the spin polarization below 
the calculated Curie temperature.} \comAK{This approach 
%In this paper we explore the electronic and magnetic properties of tetrataenite by DFT+DMFT approach. \comAK{To study electronic and dominating magnetic correlations, as well as the formation of local magnetic moments, we consider \textit{paramagnetic} phase, by artificially switching off spin polarization.
%
%Our results show 
%This 
allows us to trace the presence of well-localized magnetic moments on Fe sites, similarly earlier consideration of alpha- \cite{alpha_iron2010,OurAlphaIgoshevKatanin,OurAlphaBelozerovKatanin}, gamma- \cite{OurGamma2013}, and epsilon-iron \cite{e-iron}}. \comAK{At the same time, we find that}
%while 
magnetism of Ni sites is much more itinerant. We assume that the local magnetic moments further order magnetically due to magnetic exchange, which, as we argue, is most likely of RKKY-type.
We also find that electronic and magnetic properties of constituent Fe atoms resemble those in pure bcc Fe,
%though the ordered FeNi has a slightly tetragonally distorted fcc lattice.
%that may be associated
%
%Our results suggest that electronic properties of tetrataenite
that can be explained by peculiarities of density of Fe $3d$ states, which has pronounced peaks near the Fermi level.

\section{Method and computational details} \label{sec:computational_details}

We employ a fully charge self-consistent DFT+DMFT approach~\cite{charge_sc} implemented with plane-wave pseudopotentials~\cite{espresso,Leonov1}.
The Perdew-Burke-Ernzerhof form of 
%generalized gradient approximation 
GGA was considered.
We adopt the experimental lattice constants ${a=b=3.582}$~\AA\, and ${c=3.607}$~\AA~\cite{Kotsugi2014}, corresponding to the face-centered tetragonal (fct) latice, with four atoms per unit cell. 
Nevertheless, our calculations were performed for an equivalent but more compact unit cell of two atoms, that corresponds to a body-centered tetragonal (bct) lattice with ${a^\prime=a/\sqrt{2}}$ and ${c^\prime=c}$.
%\comAK{We have verified that the results without tetragonal distortion are qualitatively similar to those including the distortion, so that the role of the distorsion in the considered compound is mainly to introduce the magnetic anisotropy.}
%
%The convergence threshold for total energy was set to $10^{-6}$~Ry.
%
The integration in the reciprocal space was carried out using 10$\times$10$\times$10\, $\textbf{k}$-point mesh.
%
%From converged DFT results we have constructed effective Hamiltonians in the basis of Wannier functions, which were built as a projection of the original Kohn-Sham states to site-centered localized functions as described in Ref.~\cite{Korotin08}, considering $3d$, $4s$ and $4p$ states. 
%
Our DFT+DMFT calculations explicitly include the $3d$, $4s$ and $4p$ valence states, by constructing a basis set of atomic-centered Wannier functions within the energy window spanned by the $s$-$p$-$d$ band complex~\cite{Wannier}.

We perform DMFT calculations with the Hubbard parameter ${U\equiv F^0=4}$~eV, the Hund's 
%rule 
coupling ${J_{\rm H}\equiv (F^2+F^4)/14=0.9}$~eV for both Fe and Ni,
where $F^0$, $F^2$, and $F^4$ are the Slater integrals.
These values are in agreement with estimates for elemental iron~\cite{Belozerov2014}.
We also checked that considering ${U=3}$~eV for Ni as well as for both Fe and Ni %leads to similar results. 
does not qualitatively affect the results. 
%We also consider ${U=3}$~eV for Ni to make sure that it does not qualitatively affect the results. 
%
%The on-site Coulomb interaction was considered in the density-density form.
%
%The double-counting correction was taken in the fully localized limit.
We use the fully localized double-counting correction, evaluated from the self-consistently determined local occupations, to account for the electronic interactions already described by DFT.
We also verified that the around mean-field form of double-counting correction leads to similar results. %, although providing a slightly larger ($\sim$0.1) filling of \textit{d} states.
The impurity problem was solved by the hybridization expansion continuous-time quantum Monte Carlo method~\cite{CT-QMC} with the density-density form of Coulomb interaction.
To compute the density of states, we perform the analytical continuation of self-energies to real-energy range by using the Pad\'e approximants~\cite{Pade}.
\comAB{In our calculations we switch off the spin polarization by assuming spin-independent self-energy, except for calculation of uniform magnetic susceptibility, which is computed in the paramagnetic phase to extract the Curie temperature of the considered substance.}
%%%%%%%%%%%%%%%%%%%%%%%%%%%%%%%%%%%%%%%%%%%%%%%%%%%%%%%%%%
\section{Results and discussion}

\begin{figure*}[t]
\centering
\includegraphics[clip=true,width=1.0\textwidth]{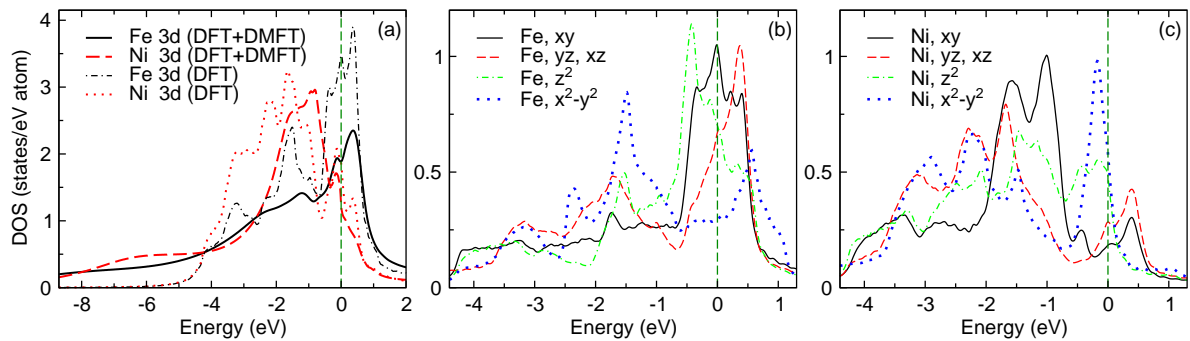}
\caption{
\label{dos}
Total (a) and orbital-projected (b,c) density of $3d$ states obtained by non-magnetic DFT (a,b,c) \comAB{calculations} \comAB{in comparison with} DFT+DMFT (a) at temperature ${T=1160}$~K. The Fermi level is at zero energy.}
\end{figure*}

First we present the density of $3d$ states (DOS) in figure~\ref{dos}(a).
Interestingly, the DOS for both constituents shows a significantly different behavior near the Fermi level
comparing to pure fcc Fe and Ni.
In particular, the DOS for Fe in FeNi has a broad peak in the vicinity of the Fermi level, with a maximum located $\sim$0.4~eV above it.
The peak resembles that in bcc Fe (see Ref.~\cite{alpha_iron2010}), 
where it originates from the $e_g$ states.
By contrast, in fcc Fe the peak is much smaller and is located below the Fermi level~\cite{OurGamma2013}. 
The van Hove singularity formed by the $t_{2g}$ states in
pure fcc Ni~\cite{Hausoel} is absent in the DOS of Ni constituent, which has only a small peak at $\sim$0.2~eV below the Fermi level.
At the same time, a satellite structure at about $-6$~eV, observed experimentally in pure Ni~\cite{satellite}, is also present in L1$_0$ FeNi.
In both materials, the DFT alone does not reproduce this %incoherent feature
satellite
originating from many-body effects.

The orbital-projected DOS presented in figure~\ref{dos}(b,c) shows that the wide peak in Fe DOS consists of contribution from all states except that of ${x^2{-}y^2}$ symmetry. 
By contrary, in Ni DOS only the ${x^2{-}y^2}$ states form a peak slightly below the Fermi level. 

\begin{figure}[t]
\centering
\includegraphics[clip=true,width=0.55\textwidth]{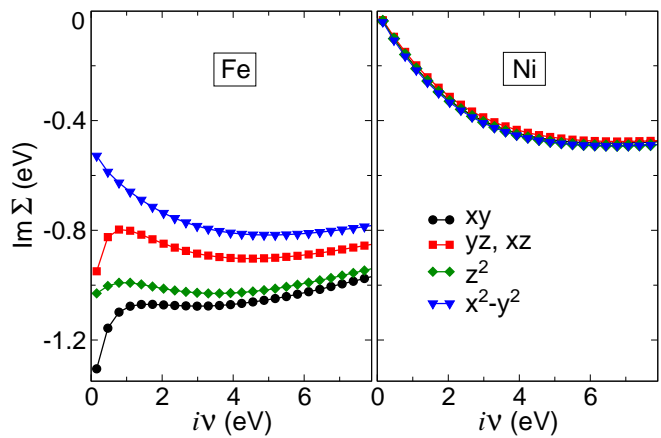}
\caption{%(Color online)
\label{Fig:self_energy}
Imaginary part of electronic self-energy on Fe (left part) and Ni (right part) sites as a function of imaginary frequency $i\nu$ obtained by DFT+DMFT at temperature ${T = 580}$~K.}
\end{figure}

In figure~\ref{Fig:self_energy} we display the imaginary part of electronic self-energy $\Sigma(i\nu_n)$ as a function of fermionic Matsubara frequency $\nu_n$.
For Fe atoms, only the self-energy for ${x^2{-}y^2}$ states has a Fermi-liquid-like form, but indicates a small life-time of quasiparticles, which is inverse proportional to $\textrm{Im}\Sigma({i\nu\rightarrow 0})$.
%has a quite large quasiparticle damping $\Gamma=$,
%(i.e., inverse quasi-particle lifetime),
At the same time, all other states show the non-coherent behavior, which is similar to that of $e_g$ states in bcc Fe~\cite{alpha_iron2010,OurAlphaIgoshevKatanin,OurAlphaBelozerovKatanin}, due to the above mentioned peaks of the DOS. 
The non-Fermi-liquid behavior is not observed in calculations with ${J_{\rm H}=0}$, that indicates an important role played by the Hund's exchange.
The behavior of self-energies for Ni atoms is completely different than that for Fe sites, namely, similarly to pure Ni \cite{Hausoel}, all $\Sigma(i\nu_n)$ have a 
Fermi-liquid form with small quasiparticle damping, indicating a presence of long-lived quasi-particles with average mass enhancement factor ${m^*/m=1.23}$.

\begin{figure}[b]
\centering
\includegraphics[clip=true,width=0.55\textwidth]{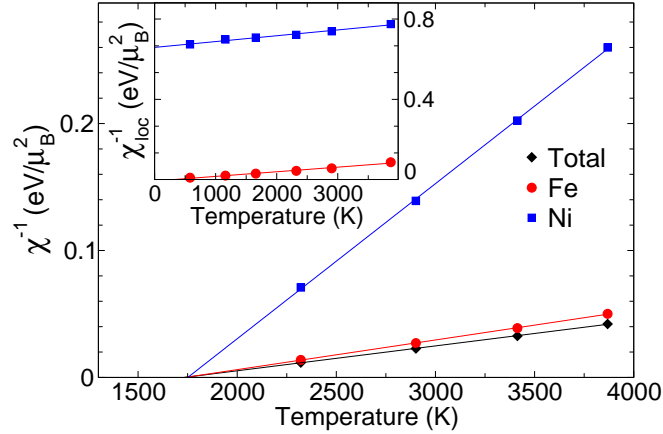}
\caption{%(Color online)
\label{Fig:chi_uniform}
Temperature dependence of inverse uniform (main panel) and local (inset) magnetic susceptibility calculated by DFT+DMFT. The straight lines depict the least-squares fit to the linear dependence.}
\end{figure}

%The calculated uniform magnetic susceptibility
Next we calculate the uniform magnetic susceptibility, which
shows Curie-Weiss behavior (see figure~\ref{Fig:chi_uniform}).
The dominant contribution to uniform susceptibility is provided by Fe sites, while the Ni sites contribution is subleading. The Curie temperature, extracted from the linear extrapolation of inverse susceptibility is  ${T_{\rm C}\approx 1750}$~K. Counting two times overestimation of the Curie temperature due to the Ising symmetry of Hund's exchange and mean-field approximation (see Refs.~\cite{Hausoel,BelozerovSU2}), we find expected Curie temperature $\sim$850~K, which agrees well with the experimental data.

\begin{figure}[b]
\centering
\includegraphics[clip=true,width=0.53\textwidth]{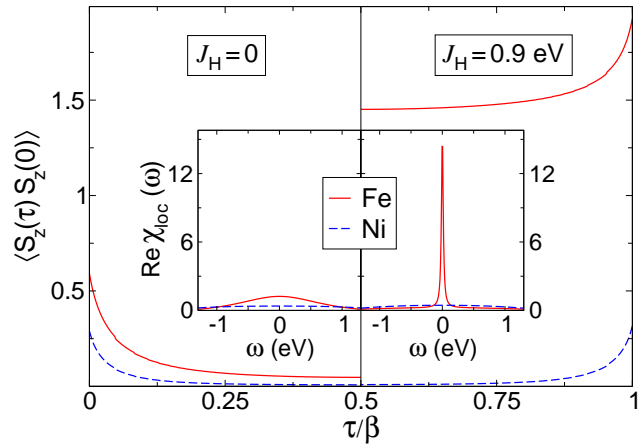}
\caption{%(Color online)
\label{Fig:chi_loc}
%Temperature dependence of inverse local (left panel) magnetic susceptibility and
Local spin-spin correlation function for Fe and Ni sites in the imaginary-time (main panel) and real-frequency (inset) domains calculated by DFT+DMFT at temperature ${T = 1160}$~K with $J_{\rm H}=0$ (left part) and $J_{\rm H}=0.9$~eV (right part).}
\end{figure}

Local static susceptibility
%\begin{equation}
$\chi_{\rm loc}=4\mu_{\rm B}^2 \int_0^\beta {\chi_{\rm loc}(\tau) d\tau},$
%\end{equation}
where $\chi_{\rm loc}(\tau) = \langle S_z(\tau) S_z(0) \rangle$,
also shows Curie-Weiss behavior (see inset of Fig.~\ref{Fig:chi_uniform}) with negative Weiss temperature $T_{\rm loc}$, which absolute value corresponds to the Kondo temperature $T_{\rm K}$ up to the numerical prefactor of order of one \cite{Wilson,Melnikov,Tsvelik}. Similarly to bcc iron, the contribution of Fe sites is characterized by small $T_{\rm K}$, showing well-formed local moments. On contrary, Ni sites are characterized by rather large Kondo temperature $T_{\rm K}$, i.e. absence of local moments, which is in line with the Fermi-liquid-like self-energies for these sites.  
 
%The imaginary time dependence of the dynamic susceptibilities ${\chi_{\rm loc}(\tau) = \langle S_z(\tau) S_z(0) \rangle}$ is shown in figure~\ref{Fig:chi_loc}, together with the real-frequency dependence of \com{real part of} $\chi_{\rm loc}(\omega)$, obtained by Fourier transform and analytical continuation using Pad\'e approximants~\cite{Pade}.
%
The imaginary time dependence of the dynamic susceptibilities ${\chi_{\rm loc}(\tau) = \langle S_z(\tau) S_z(0) \rangle}$ is shown in figure~\ref{Fig:chi_loc}, together with the real part of $\chi_{\rm loc}(\omega)$, obtained by Fourier transform and analytical continuation to real frequency~$\omega$ using Pad\'e approximants~\cite{Pade}.
Fe~sites are characterized by the finite broad plateau in the time dependence and sharp peak in frequency dependence, which confirms existence of well-defined local magnetic moments on these sites.
%
%\com{The half width of the peak in ${{\rm Re}\chi_{\rm loc}(\omega)}$ at half of its height yields approximately inverse lifetime of local magnetic moments~\cite{OurGamma2013}.}
%
The local moments disappear with switching off Hund's exchange, that shows explicitly that the obtained peculiarities of magnetic properties of Fe sites are the characteristics of Hund's metal behavior~\cite{Hund_metals}.
At the same time, Ni sites have almost vanishing correlation function at finite imaginary times ${\tau\sim\beta/2}$, and absent peak of the frequency dependence, independently of Hund's exchange, showing once more the absence of local moments on these sites. 

Finally we consider the particle-hole bubble (irreducible static non-uniform magnetic susceptibility) 
\begin{eqnarray}
\chi_{\bf q}^{0} &=& -\frac{2\mu_B^2}{\beta} \sum_{{\bf k},\nu_n,i,j,m,m'} G^{im,jm'}_{\bf k} (i\nu_n) G^{jm',im}_{\bf k+q}(i\nu_n),
\end{eqnarray}
where $G^{im,jm'}_{\bf k} (i\nu_n)$ is the one-particle Green function for $d$-states obtained using the Wannier-projected Hamiltonian, $i,j$ and $m,m'$ are the site and orbital indexes, respectively.
%, $\nu_n$ are the fermionic Matsubara frequencies.

\begin{figure}[t]
\centering
\includegraphics[clip=true,width=0.51\textwidth]{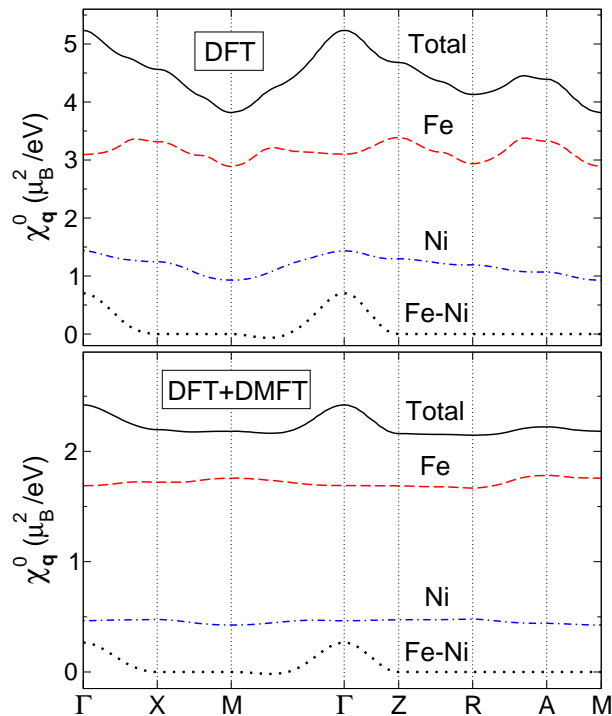}
\caption{\label{fig:chi_q}
Momentum-dependence of the particle-hole bubble
%irreducible magnetic susceptibility
and its partial contributions from Fe and Ni sites obtained within DFT (top panel) and DFT+DMFT (bottom panel) at temperature ${T = 1160}$~K.}
\end{figure}

One can see (Fig.~\ref{fig:chi_q}) that $\chi_{\bf q}^{0}$ has its maximum at $\Gamma$ point (${{\bf q}=}0$), implying that ferromagnetism is favoured in both DFT and DMFT approaches. However, this maximum appears mainly because of the
%contribution of  Ni sites and
mixed Fe-Ni contributions. At the same time, the contribution of iron sites does not favour ferromagnetism in DFT, and is almost momentum-independent in DMFT, similarly to bcc iron \cite{OurAlphaIgoshevKatanin}. For L1$_0$ FeNi the latter weak momentum dependence occurs since electronic states, corresponding to major part of Fe orbitals (except ${x^2{-}y^2}$), show the non-Fermi liquid behavior, and the main contribution to non-uniform susceptibility in DMFT is provided by mixed Fe-Ni part. This is similar to the mixed $t_{2g}$-$e_g$ contribution to particle-hole bubble in bcc iron \cite{OurAlphaIgoshevKatanin,OurAlphaBelozerovKatanin}, with $t_{2g}$ states being more itinerant, while $e_g$ states being more localized.
Since momentum dependence of particle-hole bubble can be related to the magnetic exchange \cite{OurAlphaIgoshevKatanin,OurAlphaBelozerovKatanin}, from these findings one can expect RKKY-type of magnetic exchange in L1$_0$~FeNi, with the states corresponding to nickel sites (including virtual hopping to iron sites)  providing itinerant contribution and iron states providing local moment contribution.

\comAB{We have verified that the results without tetragonal distortion are qualitatively similar to those including it, so that the role of the distortion in the considered compound is mainly to introduce the magnetic anisotropy.}

\section{Conclusion} \label{sec:conclusions}

Our DFT+DMFT study of L1$_0$ FeNi has revealed that magnetism of Fe and Ni sites is completely different. The former is characterized by well-defined local magnetic moments, which appear due to Hund's exchange,
%which are similar to those in pure bcc Fe, % though the lattice of ordered FeNi is 
while the latter is much more itinerant.
Although the crystal structure is close to the fcc one,
the localized magnetism and behavior of self-energy of Fe sites resemble those in pure bcc Fe.
This can be explained by presence of peaks in the DOS near the Fermi level in L1$_0$ FeNi, similar to that in $e_g$ states of bcc Fe.

The estimated Curie temperature agrees with the experimental data, up to the factor of two, which appears because of considering Ising symmetry of Hund's exchange and limitations of dynamical mean-field theory.
The high Curie temperature of tetrataenite can be explained by large DOS at the Fermi level, which, according to the Stoner criterion, leads to the strong ferromagnetic instability. In this respect, tetrataenite is quite similar to elemental (bcc) iron, also having large DOS at the Fermi level due to nearby peaks.

The calculated local magnetic susceptibility confirms different magnetic behavior of Fe and Ni sites. The obtained momentum dependence of particle-hole bubble suggests that the magnetic exchange in L1$_0$ FeNi is of RKKY-type, with iron states providing local-moment contribution, and the states corresponding to nickel sites (including virtual hopping to iron sites) providing itinerant contribution.

%\com{AK: Is this really needed here for Conclusion? I have inserted short phrase above about Hund exchange, which seems to me sufficient.
%Our results indicate that the Hund's exchange in L1$_0$ FeNi is a major source of significant electronic correlations on Fe sites, leading to formation of local magnetic moments. In this respect, L1$_0$ FeNi can be considered as a site-selective Hund's metal, in which the electronic degrees of freedom are driven by Hund's exchange for only a part of atoms.}

Therefore, L1$_0$ FeNi is a good candidate for rare-earth-free magnets. Further studies (e.g., its doping/alloying) may find the way for its better phase stability. The 
search of other rare-earth-free magnets will also allow one to achieve 
%higher coercive force \com{(AB: $\rightarrow$ 
better performance characteristics, such as coercive force and magnetic energy product.
%and better phase stability. 
Both theoretical and experimental investigations of such magnets have to be performed.

% provides a better understanding which can improve the laboratoy systesis and help in search of new high-performance permanent magnets.
% and also for substitution materials with intermediate performance and lower prices

%\vspace{-0.5cm}
\section*{Acknowledgments}
%\begin{acknowledgments}
%\vspace{-0.5cm}
The DMFT calculations were supported by the Russian Science Foundation (project 19-72-30043).
The DFT calculations were supported by the Ministry of Science and Higher Education of the Russian Federation (theme “Electron” No. AAAA-A18-118020190098-5).

%end{acknowledgments}

\end{document}